\documentclass[useAMS,usenatbib]{mn2e}

\def\M{M$_{\odot}$}
\def\R{R$_{\odot}$}

 \def\Mej{$M_{\rm ej}$}

\usepackage{graphicx, subfigure}
\usepackage[authoryear]{natbib}
\usepackage{placeins}
\usepackage{amsmath}

\title[Double-peaked SLSNe]
{Seeing double: the frequency and detectability of double-peaked superluminous supernova light curves}

\author[Nicholl et al.]{M. Nicholl$^{1,2}$\thanks{E-mail: matt.nicholl@cfa.harvard.edu} and
S. J. Smartt$^{1}$
\\
$^1$  Astrophysics Research Centre, School of Mathematics and Physics, Queen's University Belfast, 
Belfast, BT7 1NN,  UK \\
$^2$  Harvard-Smithsonian Center for Astrophysics, 60 Garden Street, Cambridge, Massachusetts 02138, USA \\
}

\begin{document}

\maketitle

\begin{abstract}

The discovery of double-peaked light curves in some superluminous supernovae offers an important new clue to their origins. We examine the published photometry of all Type Ic SLSNe, finding 14 objects with constraining data or limits around the time of explosion. Of these, 8 (including the already identified SN 2006oz and LSQ14bdq) show plausible flux excess at the earliest epochs, which deviate by 2-9$\sigma$ from polynomial fits to the rising light curves. Simple scaling of the LSQ14bdq data show that they are all consistent with a similar double-peaked structure. PS1-10pm provides multicolour UV data indicating a temperature of $T_{\rm bb}=25000\pm5000$\,K during the early `bump' phase. We find that a double-peak cannot be excluded in any of the other 6 objects, and that this behaviour may be ubiquitous. The homogeneity of the observed bumps is unexpected for interaction-powered models. Engine-powered models can explain the observations if all progenitors have extended radii or the central engine drives shock breakout emission several days after the supernova explosion.

\end{abstract}


\section{Introduction} \label{intro}

Superluminous supernovae (SLSNe) are explosions with absolute magnitudes $M<-21$ \citep{qui2011,gal2012}. Most are hydrogen-poor, and have been termed SLSNe I or Ic. A combination of 
relative scarcity \citep{qui2013,mcc2015} and a preference for faint galaxies \citep{nei2011,chen2013,lun2014,lel2015} meant that these objects went undiscovered (or unrecognised) for many years. 
Unbiased surveys, such as the Palomar Transient Factory \citep[PTF;][]{rau2009},  Catalina Real Time Transient Survey \citep{dra2009}, Pan-STARRS \citep{kai2010}, and La Silla QUEST \citep[LSQ;][]{balt2013} have fed spectroscopic follow-up to quantify their energy and origins. PTF had a built-in spectroscopic component \citep[e.g.][]{vre2014}, Pan-STARRS had significant 8m time for high-$z$ follow-up
\citep[e.g.][]{lun2013}, and LSQ feeds the Public ESO Spectrosopic Survey of Transient Objects 
\citep[PESSTO;][]{sma2015}, which targets SLSNe for focused study. 

The light curves can be described using models powered by a central engine, such as a nascent millisecond magnetar \citep{kas2010,woo2010}, or by thermalisation of the ejecta energy in a massive circumstellar medium \citep{che2011,gin2012}. Distinguishing between these two has proved problematic despite quantitative fitting of the bolometric light curves and extensive spectra \citep[e.g.][]{ins2013}. %Pair-instability explosions are also a plausible explanation for some SLSNe \citep{gal2009}, though this interpretation is disputed \citep{des2012,nic2013}.

A clue lies in the early light curve. \citet{lel2012} detected a fast initial peak, or `bump', in the \textit{ugriz} light curve of SN 2006oz, preceding a 25-30 day rise to peak. \citet{nic2015a} observed a similar bump, in one filter but with better time resolution, for the slowly rising LSQ14bdq. Both were interpreted as cooling emission in either CSM \citep{lel2012} or extended stellar material \citep{nic2015a,piro2015}, with an implied radius of $>500$\,\R. \cite{kas2015} proposed that it could  be the signature of a second shock breakout, driven by a central engine in the expanding ejecta. Regardless of the mechanism, the existence of bumps in both fast and slow SLSNe is particularly interesting given that it is still unclear whether these objects form a continuum or two distinct classes \citep{nic2015b}. If such bumps were ubiquitous, it would help to constrain the origin of these explosions and the progenitor structure. 

This paper aims to establish whether these two bumps were special cases, or if they may in fact be common -- and frequently missed by the current generation of SN surveys. 

%% TO SAVE SPACE, IN  A LETTER, WE CAN SKIP THE PAPER DESCRIPTION 
%%
%In Section \ref{bumps}, we show objects with possible bump detections. In Section \ref{nobumps}, we look at SLSNe that instead have deep pre-discovery limits at potentially constraining epochs. We attempt to derive some properties of the bump, and examine future detectability, in Section \ref{props}, before concluding in Section \ref{conc}.

\begin{table}
\caption{SLSNe with early flux excesses or constraining limits}
\label{tab}
\begin{tabular}{lcccc}
\hline
Name & $z$ & Significance ($\sigma$)$^a$ & $t$-stretch$^b$ & $m$-stretch$^c$ \\
\hline
\hline
LSQ14bdq	& 0.345 &	9.2, 15.8, 9.3,&	--	&	--	\\
                        &   &   10.0, 10.7      &             &           \\	 
SN 2006oz	& 0.376 & 10.8, 18.3				&	0.43	&	1	\\
\hline
PTF09cnd	& 0.258 &4.3					&	0.83	&	1	\\
PS1-10pm	& 1.206 &	5.5, 6.8, 3.8				&	0.50	&	0.67	\\
PS1-10ahf	& 1.158 &	4.1		&	0.57	&	0.50	\\
iPTF13ajg	& 0.740 &	3.6		&	0.63		&	0.48	\\
SNLS06D4eu	& 1.588 &	9.3		&	0.36	&	0.67	\\
SN1000+0216	& 3.899 &	5.0, 2.7	&	1	&	0.67	\\
\hline
SN 2011ke	& 0.143 &	--	&	0.4	&	-- \\
LSQ12dlf		& 0.255 &	--	&	0.7	&	-- \\	
SCP06F6		& 1.189 &	--	&	0.85	&	-- \\
SNLS07D2bv	& 1.50 &	--	&	0.65	&	-- \\
PS1-10awh	& 0.908 &	--	&	0.62	&	-- \\
PS1-10bzj	& 0.650 &	--	&	0.45	&	-- \\
\hline
\end{tabular}
$^a$Deviation of the early excess (chronological);
$^b$Time stretch-factor to match LSQ14bdq light curve to data;
$^c$Magnitude stretch-factor to match LSQ14bdq.
\end{table}

\section{SLSNe with plausible bumps}\label{bumps}

We have examined the early light curves of all published SLSNe Ic as of August 2015:  
see \cite{nic2015b} and references therein; and of particular interest are objects in 
\citet{qui2011}; \citet{mcc2015}; \citet{how2013}; \citet{coo2012}; \citet{vre2014}. 
To be useful for our study, detections or limits must exist $>25$\,d before maximum light, in the case of fast-evolving SLSNe like SN 2006oz, or $>60$\,d for slow-risers like LSQ14bdq.

\begin{figure}
\centering
\includegraphics[width=8.25cm,angle=0]{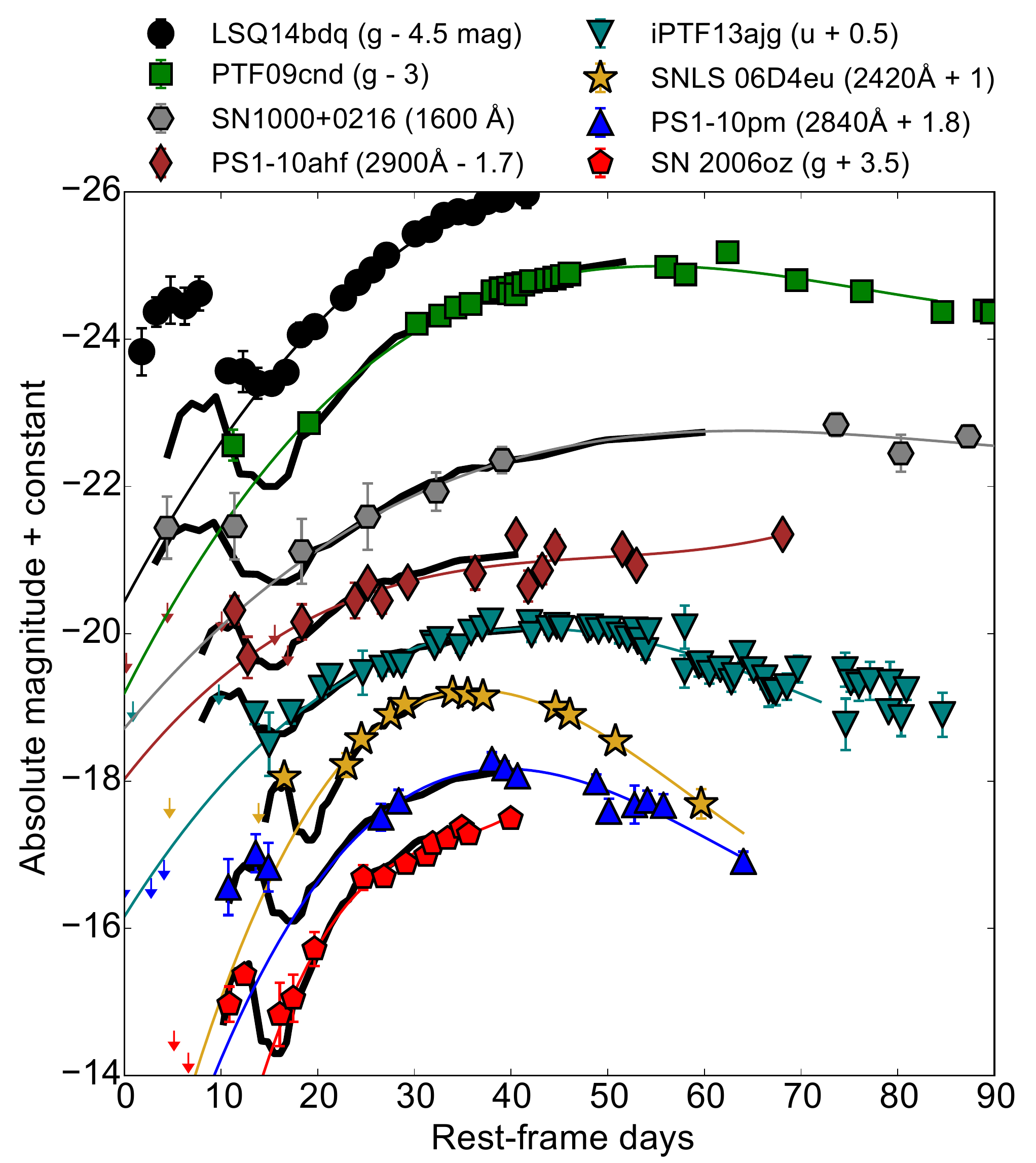}
\includegraphics[width=8.25cm,angle=0]{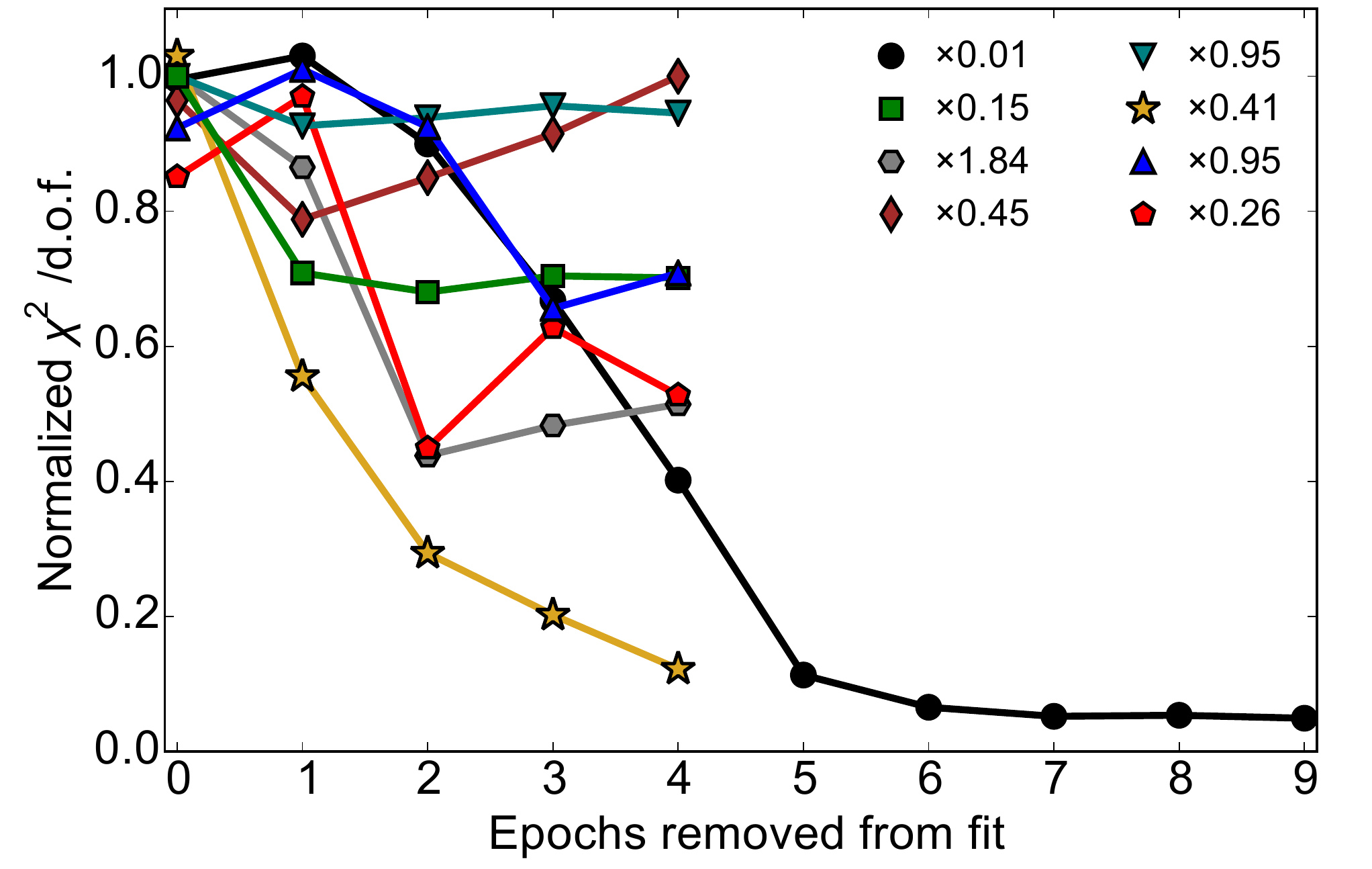}
\caption{
Top: Observed data and polynomial fits to the rising light curves (thin coloured lines). The thick black lines are the scaled LSQ14bdq light curves. 
Bottom: The reduced $\chi^2$ for polynomial fits as early points are successively excluded from polynomial fitting. The polynomials in the top panel are those for which $\chi^2$/d.o.f.~shows the steepest drop. Details and data sources are given in Section \ref{bumps}. 
 }\label{fig:bumps}
\end{figure}

We identify six additional SLSNe that show non-monotonic behaviour or flux excesses at the earliest epochs. Their light curves are plotted in Figure \ref{fig:bumps}. Of the objects shown, only LSQ14bdq and SN 2006oz were previously recognised as double-peaked SLSNe, although \citet{how2013} did note a flux excess in SNLS06D4eu at early times. For the other objects, such excesses were presumably dismissed as photometric noise. However, when presented together with similar objects, it seems plausible that the earliest data points indicate a real (under-sampled) bump.

We fit the light curves (up to 30 days after maximum) with third-order polynomials and then successively remove the earliest points and re-fit, looking at the effect on $\chi^2$ per degree of freedom (d.o.f.). When this value shows its largest decrease, we assume that we have removed the early excess from our fit. We then measure the deviation of the excluded points from this `bump-free' polynomial fit, in units of $\sigma$ (the quoted photometric error; Table\,\ref{tab}).
Additionally, we scale and overlay the light curve of LSQ14bdq (the best sampled bump detection) to test if the shape of the excess in each event is consistent with a double-peak of similar structure. The scaled light curve is not fit to the data in any formal sense -- it is simply the best visual match. 
If data are in a similar rest-frame filter to LSQ14bdq ($g$-band), we allow a stretch factor in time ($t'=t\times t_{\rm stretch}$) and an additive shift in magnitude. For LSQ14bdq, the difference in $g$-band magnitude between the bump and main peak is $M_{\rm bump}-M_{\rm peak}=2$\,mag. For higher-$z$ SLSNe with rest-frame UV photometry, the early excess (if interpreted as a bump) appears to 
indicate $M_{\rm bump}-M_{\rm peak}<2$\,mag. 
In such cases, we also apply a scaling factor in magnitude ($M'=M\times M_{\rm stretch}$); $M_{\rm stretch}<1$ flattens the light curve, reducing the brightness ratio between the two peaks.

{\bf LSQ14bdq and SN 2006oz:} 
The two objects previously identified as double-peaked 
show marked changes in $\chi^2$/d.o.f., and early excesses $>9\sigma$ compared to the polynomial fits, demonstrating the utility of our method.
The light curve of LSQ14bdq can be mapped almost perfectly onto SN 2006oz using only a stretch in time (and simple shift in brightness).  LSQ14bdq is slowly-evolving, whereas SN2006oz is a much faster SLSN \citep[see][for discussion of the physical interpretation]{nic2015b}. The simultaneous match to both peaks with a single stretch factor suggests that the peak widths may be correlated. The excellent correspondence of these two objects gives us confidence that we can assume LSQ14bdq is representative of double-peaked SLSNe, up to some stretch factor.

\begin{figure}
\centering
\includegraphics[width=8.25cm,angle=0]{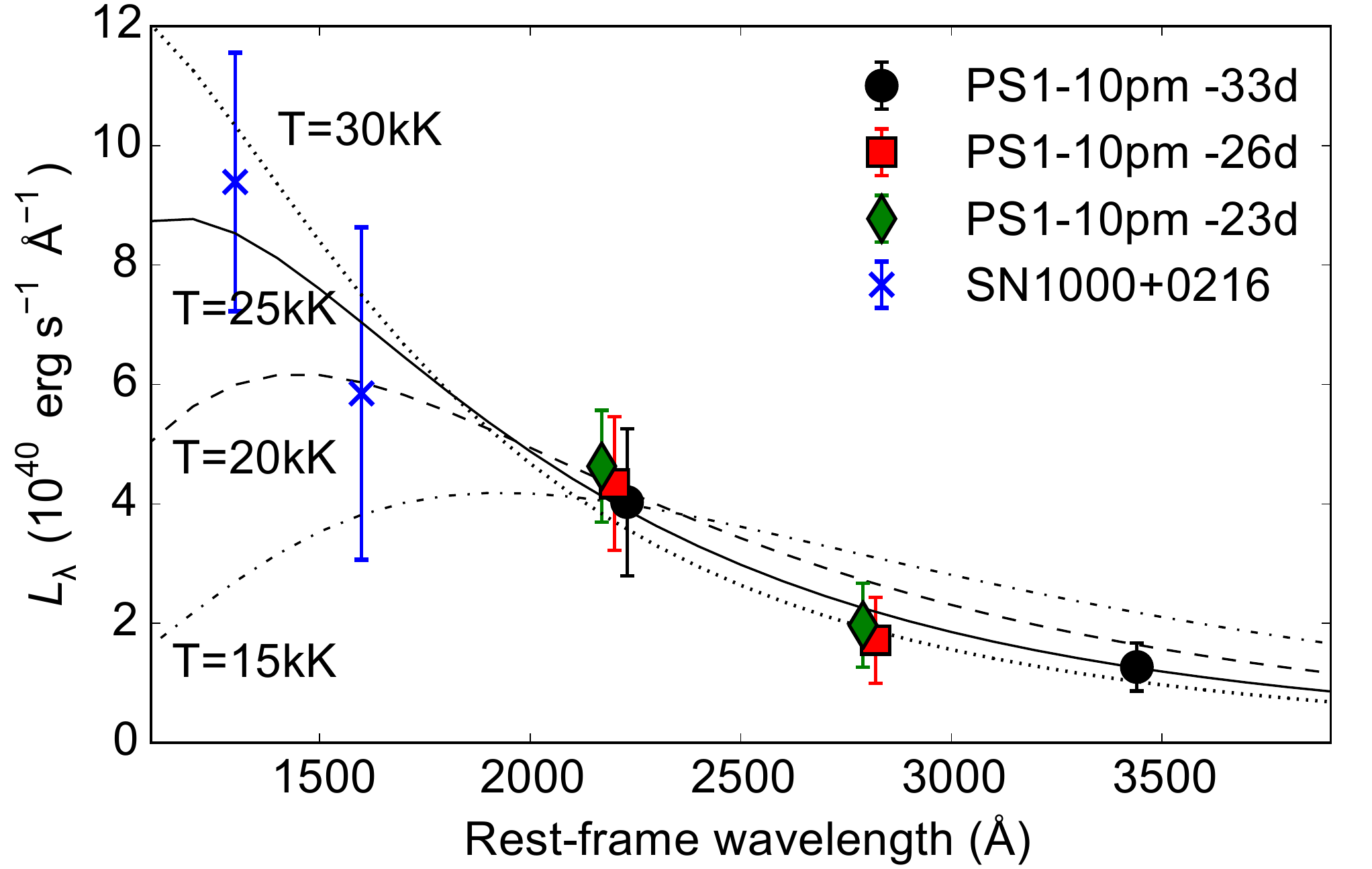}
\caption{The SED of PS1-10pm compared to blackbody curves. 
 For illustration, the blue asterisks give the FUV luminosity of 
SN1000+0216 during its initial peak, at around $-50$\,d.}\label{fig:sed}
\end{figure}

{\bf PS1-10pm:} The PS1-10pm data from \citet{mcc2015} shows flux excess in the three earliest $r_{\rm P1}$ points (2800\,\AA\ rest-frame for $z=1.206$)  with a deviation of $>4\sigma$. 
This was not recognised as real structure, 
and one could still reasonably assume a simple broad rise (albeit shallower than the subsequent decline).
However, the scaled light curve of LSQ14bdq is an excellent match to the early points and rising phase. Moreover, PS1-10pm also showed significant excess in the $g_{\rm P1}$ and $i_{\rm P1}$ filters \citep{mcc2015}. The multiple UV data points allow a blackbody SED fit and temperature estimation. Fluxes 
were corrected for redshift, luminosity distance and Milky Way foreground
extinction. No internal galaxy extinction has been applied, since this is unknown (any host reddening would serve to increase the temperature of the fit). The SED is consistent with a blackbody of $T_{\rm bb}=25000\pm5000$\,K (Figure\,\ref{fig:sed}), giving the first UV measurement of the bump temperature. This is similar to the temperature of $T_{\rm bb} = 15000 \pm 5000$\,K implied by the optical data for SN 2006oz \citep{lel2012}.

{\bf SNLS06D4eu:}
 \citet{how2013} already noted that SNLS06D4eu showed a likely early-time excess in filters with effective wavelengths between $\sim 2400$-3500\,\AA~in the rest-frame. We confirm this with both the significance 
of the deviation from the polynomial fit ($9.3\sigma$), and a convincing match with the LSQ14bdq template, indicating that the earliest point is quite consistent with an LSQ14bdq-like bump.

{\bf PTF09cnd:}
Excluding the first point from the polynomial fit 
gives a clear decrease in $\chi2$/d.o.f., and indicates an excess of $4.3\sigma$.
It can be well matched by LSQ14bdq, which it also resembles spectroscopically \citep{nic2015a,qui2011}. We consider this a likely detection.

{\bf PS1-10ahf:}
Although the data are noisy \citep{mcc2015}, 
removing the first point gives a minimum in $\chi^2$/d.o.f.; the excess is then $4.1\sigma$.
The UV light curve rise is shallow, and the optical LSQ14bdq data require a significant stretch in magnitude to match it. However, the overall shape, and detection limits, are consistent with a bump.

{\bf iPTF13ajg:}
This SLSN, from \citet{vre2014}, 
shows a marginal detection at best, with only a small variation in $\chi^2$. On the other hand, the deviation of the first point is $3.6\sigma$ and the shape is consistent with LSQ14bdq.

{\bf SN 1000+0216:}
This object is the most distant SLSN yet detected. 
It was not spectroscopically classified, but we include it here because of an apparent flux excess
in the rebinned $i$-band light curve from \citet{coo2012}. This samples rest-frame emission at $\approx1600$\,\AA. The $\chi^2$ test shows a sharp drop, and the resultant deviation is $5.0\sigma$. The shape is similar to the $g$-band light curve of LSQ14bdq, with no time stretch required. A magnitude stretch of 0.67 is needed, similar to factors for other UV SLSNe (Table \ref{tab}). Overall, the flux differences between the first and second peaks are lower in the UV ($\sim1$-1.3\,mag) than in the optical ($\sim2$\,mag). This could mean that the initial peak has a hotter effective temperature than the second. As a consistency check, we plot the rest-frame 1300-1600\,\AA\ flux of the bump of SN1000+0216 alongside the data for PS1-10pm (correcting the 1300\,\AA\ flux for Lyman-$\alpha$ absorption along the line of sight; J.~Cooke, private communication). If we assume that these two SLSNe can
be compared, the flux is consistent with the blackbody temperature of PS1-10pm, found to be $T_{\rm bb}=25000\pm5000$\,K in Figure \ref{fig:sed}.

\begin{figure*}
\includegraphics[width=15.8cm,angle=0]{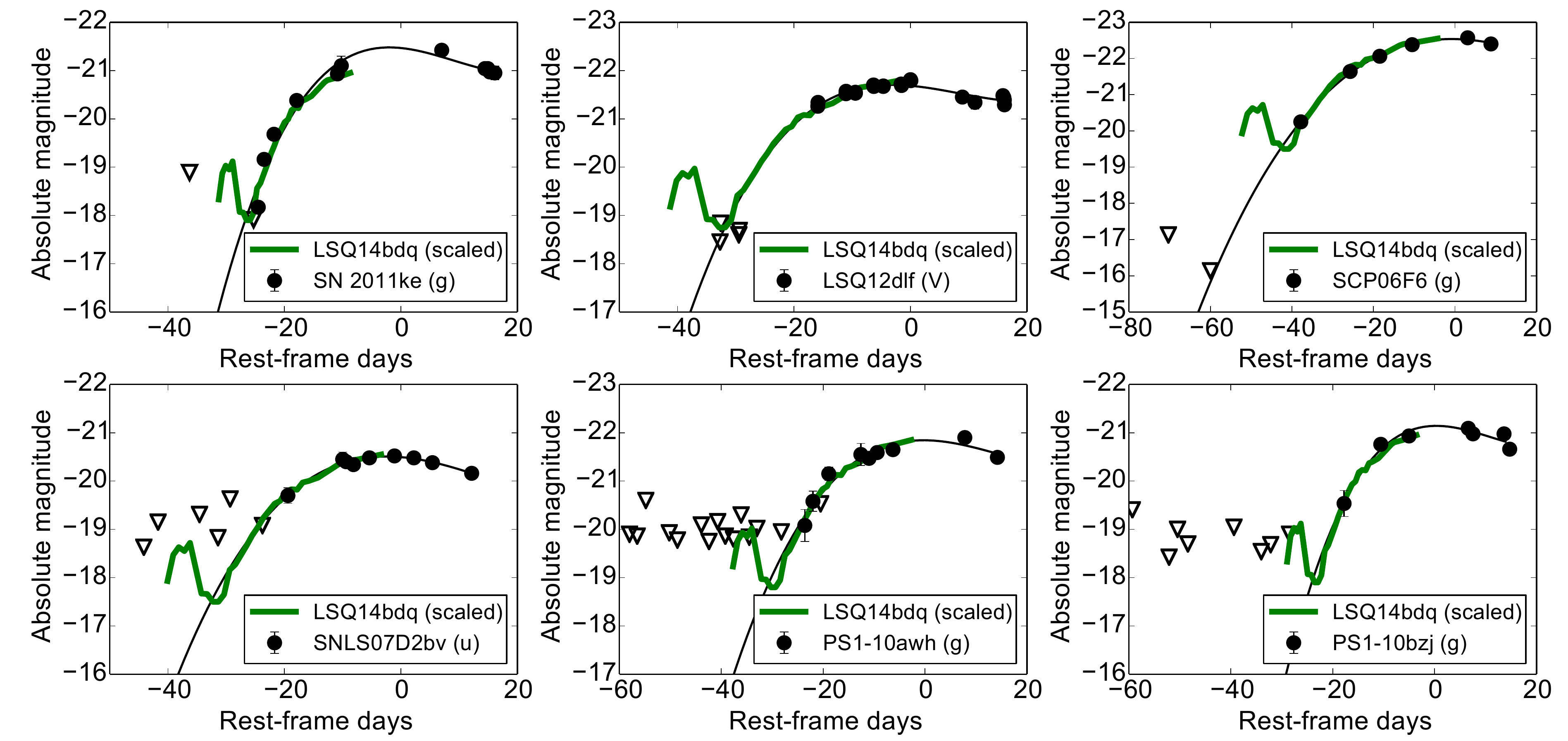}
\caption{SLSNe with limiting magnitudes at $\ga25$\,d before observed maximum. The filters indicated are in rest-frame. In the top row, the cadence is insufficient to exclude a precursor peak. In the bottom row, the limits are too shallow (PS1-10bzj is marginal).}\label{fig:limits}
\end{figure*}

\section{SLSNe with no observed bump}\label{nobumps}

We next consider SLSNe Ic with well-observed rising phases that do not show signs of non-monotonic behaviour, but that have deep pre-detection limits to potentially exclude a bump. To act as a useful constraint, the limits must exist $\ga25$ days before maximum light. Of equal importance is their depth; as LSQ14bdq and SN 2006oz exhibited a difference of $\approx2$\,mag between the bump and main peak, limits must be at least this much fainter than the main peak to exclude a similar bump. Six SLSNe Ic fulfil these criteria, and are shown in Figure \ref{fig:limits}. These are SN 2011ke \citep{ins2013}, LSQ12dlf \citep{nic2014}, SCP06F6 \citep{bar2009}, SNLS07D2bv \citep{how2013}, PS1-10awh \citep{chom2011} and PS1-10bzj \citep{lun2013}.

Following  the method employed in Section \ref{bumps}, we fit the rise phase of each SLSN with a third-order polynomial. We then scale the light curve of LSQ14bdq to match the polynomial. However, unlike in the previous section, we do not have information on $M_{\rm bump}-M_{\rm peak}$, and therefore any scaling factor in magnitude is unconstrained. Our approach here is to use photometry in a filter as close as possible to rest-frame $g$-band, and scale LSQ14bdq only along the time axis, thus implicitly assuming that $M_{\rm bump}-M_{\rm peak}=2$\,mag for each object. This is reasonable given the similarity between LSQ14bdq and SN 2006oz. On the other hand, since well-observed optical bumps exist for only these two objects, it is unclear what the true scatter in magnitude is.

For SN2011ke and LSQ12dlf, the observational cadence is low and the predicted bump falls at an epoch where there are insufficient data points.
In the case of LSQ12dlf, the limits suggest that, if there is a bump, it may be fainter than LSQ14bdq, or have a deeper `dip' between the peaks. However, earlier limits are needed to rule this out.
The detection limits for the higher-$z$ SLSNe are not quite deep enough to rule out the existence of a bump. There are 
no other published SLSNe Ic which have enough observational data 
at the early times illustrated in Figures \ref{fig:bumps} and \ref{fig:limits}. 

In summary, we cannot exclude a the existence of a bump for {\em any object} to date. Where data exist that are both deep enough, and have sufficient time sampling, we see an observational signature of a bump with 
significance between 2 - 9$\sigma$ (compared to the observational errors). 
The corollary to this statement is that the current published sample of SLSNe Ic is consistent with \emph{all} objects having a bump.

\section{Physical interpretation and implications}\label{props}

\cite{nic2015b} proposed that the bump in LSQ14bdq was post-shock cooling of extended stellar material \citep{rab2011}. Its luminosity could be recovered with reasonable explosion energy ($E_{\rm k} \sim 2\times10^{52}$\,erg) if the progenitor had radius 
$R_{\rm \star} \sim 500$\,\R\ and ejected \Mej\,$\sim30$\,\M. \cite{piro2015} then modelled the data using shock breakout in a different structure \cite[employing][]{nak2014}. 
His preferred progenitor consisted of a compact core of $M_{\rm c}\simeq30$\,\M, and a low mass envelope of $M_{\rm e}\simeq0.3$\,\M\ extending to 500-5000\,\R. This model reproduced the rapid drop-off after the bump, giving a better fit to the full 10-day light curve than did the \citet{rab2011} models.
He found a similar explosion energy of 
 $E_{\rm k} \simeq 10^{52}$\,erg. 
The underlying physics of the two models is similar, requiring 
either an inflated helium/carbon-oxygen star (since no hydrogen is detected) or low-mass envelope, produced by pre-explosion activity. If bumps exist in most SLSNe, it would imply that they all have similar progenitor structure -- somewhat unexpected for carbon-oxygen cores or helium stars.

In Section\,\ref{bumps} we fitted a blackbody to the UV flux of PS1-10pm, finding
$T_{\rm bb} =25000\pm5000$\,K.
This is consistent with the temperature of the first peak expected by 
\cite{piro2015} of 10000-20000\,K. Following \citet{nic2015a}, we fit the PS1-10pm data shown in Figure \ref{fig:bumps} with the shock cooling model of \citet{rab2011}. Assuming a blackbody SED, we apply synthetic photometry to the model in the \textit{HST} F250W filter, which has a similar effective wavelength to the data. A good match is obtained for very similar parameters to LSQ14bdq: $R_*=500$\,\R, $M_{\rm ej} \approx 30$\,\M, $E_{\rm k} \approx 2 \times 10^{52}$\,erg. The temperature around the peak of the bump is predicted to be $\sim20000$\,K, in good agreement with the data (Figure \ref{fig:sed}) and the predictions of \citet{piro2015}.

The radius associated with the 25000\,K blackbody in Figure \ref{fig:sed} is $\simeq10^4$\,\R. However, as the first detection is likely several days after explosion, this is not the radius of the progenitor. It could represent CSM or the extended material modelled by \citet{piro2015}, but is also consistent with the \citet{rab2011} fit after a few days of expansion.

%{\bf MATT : although 10pm is sparse, can you roughly fit it with your analysis and see what parameters you recover. You can fit the stretched 14bdq LC, which will then go through the 10pm data points. You should also then be able to estimate $L$,  assuming a blackbody, and $T=20$K. This will give radius $R$ of emitting surface. This should give a handle on $v$, just from $R(t)=R_e + v_e t$ (eqn 8 of Piro).  }

\cite{kas2015} proposed a different model, where ejecta from a compact progenitor expand to about 10$^4$\,\R\ after 5-10 days. If a central engine is formed (magnetar or accreting black hole) it may dynamically inflate a high-pressure bubble, propagating as 
a second shock. Evidence for this shock has been found in the flat velocity curves of SLSNe Ic \citep{chom2011,nic2015b}. The shock breaks out from the expanding ejecta, giving an optical/UV burst lasting several days. The
predicted temperature is $T_{\rm bb} \approx 20000$\,K,
the spectrum relatively featureless and
blackbody-like, and there is a range of luminosities and timescales depending on the explosion/engine energies and thermalisation efficiency. 
This temperature (and radius) are in agreement with our
measurements for PS1-10pm, and we observe a range of bump durations and luminosities (manifested in stretch factors applied to LSQ14bdq). However, in their analytic model a distinct bump is predicted only for large masses or explosion energies, or inefficient heating by the engine at early times. If bumps are common, this could be problematic for the model. Alternatively, it could place important constraints on how the magnetar energy thermalizes in the ejecta. \cite{kas2015} note that detailed hydrodynamical calculations are needed to investigate this further.

The time-stretch parameter used to scale the LSQ14bdq light curve preserves the relative widths of the bump and main peak. If this simple stretch can accommodate any SLSN light curve, it suggests that the widths of the two peaks are correlated. This could be an important discriminant between models. It is not obvious how the extended material model would produce this correlation. It may be possible to construct a model in which shock breakout occurs in circumstellar material \citep{ofek2010}, before further interaction powers a second peak. In this case, properties of the CSM would determine the duration of both peaks, though the required structure may be rather contrived. In the \citet{kas2010} model, the duration of the bump is linked to the diffusion and magnetar spin-down times, which also determine the time taken to rise to the main light curve peak, offering a possible explanation for the width relationship between the peaks.

\begin{figure}
\centering
\includegraphics[width=8.25cm,angle=0]{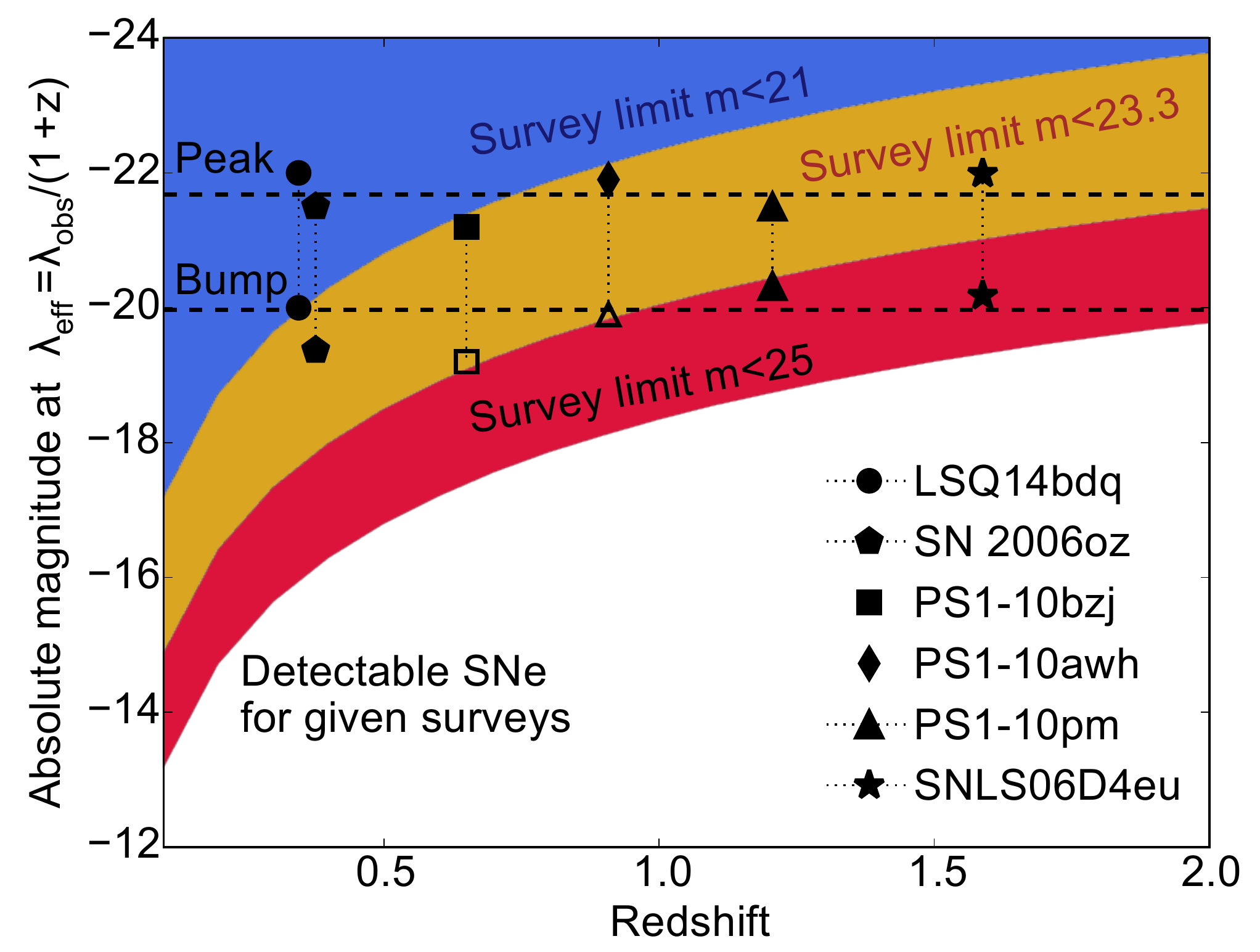}
\caption{Detectability of initial and main peaks as a function of redshift. Dashed lines give  mean magnitudes of observed peaks and bumps. Empty symbols are predicted bumps, assuming $M_{\rm bump}-M_{\rm peak}=2$\,mag. The shaded areas show what can be observed by shallow (PTF, LSQ, PS1 3$\pi$), medium-deep (PS1 MDS, SDSS) and deep surveys (SNLS, DES). PS1-10bzj  and PS1-10awh lie close to the PS1 MDS detectability threshold.}\label{fig:detect}
\end{figure}

\section{Conclusions}\label{conc}
There are 14 SLSNe Ic with published early 
photometry that  
constrains the existence of double-peaked light curves. In 8 of these, we have found at least some
evidence that a bump may exist at
blue and UV rest-frame wavelengths. 
The data for the other 6 are either too shallow or sparsely sampled 
to exclude bumps. We therefore propose that bumps may be ubiquitous, and that simple stretch factors map the well-sampled structure of LSQ14bdq onto all of them. 

An appealing, unifying explanation is that of 
\cite{kas2015}: an engine-driven shock in pre-expanded ejecta. The alternatives are the radially extended progenitor star models of \cite{nic2015a} and \cite{piro2015}, and the dense CSM interaction model of \cite{mor2012}. CSM interaction models require a double-shell structure to reproduce the fast initial rise \citep{mor2012,nic2015a}; it seems this would need to be remarkably 
homogeneous in mass, density and radius across SLSN Ic progenitors. 
The temperature we measure for PS1-10pm ($T_{\rm bb}\simeq$25000\,K), supported by far-UV data for SN1000+0216, is consistent with both \citet{kas2015} and \citet{piro2015}. 

Many SLSNe have either been at 
too high redshift or did not have the cadence to detect the precursor peaks. 
Figure \ref{fig:detect} illustrates the detectability of bumps by  
currently running surveys, illustrating that the Dark Energy Survey (DES) 
\citep{pap2015} has excellent potential to consistently detect bumps. For a typical $z\sim1$, DES \emph{griz} filters should anchor the rest-frame SED at $\sim2400$-4500\,\AA. The next step will be to gather high-cadence early spectra.

~\\
\noindent{\bf ACKNOWLEDGMENTS}{
We thank Edo Berger, Jeff Cooke, Dan Kasen, Brian Metzger, Bob Nichol and Tony Piro for helpful discussion. Funded by the European Research Council: EU(FP7/2007-2013) Grant n$^{\rm o}$ [291222].}

\bibliographystyle{mn2e}
\bibliography{bib_mn20151203}

\begin{thebibliography}{35}
\expandafter\ifx\csname natexlab\endcsname\relax\def\natexlab#1{#1}\fi

\bibitem[{Baltay {et~al}\mbox{.}(2013)Baltay, Rabinowitz, Hadjiyska, Walker,
  Nugent, Coppi, Ellman, Feindt, McKinnon, Horowitz, {et~al.}}]{balt2013}
Baltay C. {et~al.}, 2013, PASP, 125, 683

\bibitem[{Barbary {et~al}\mbox{.}(2009)Barbary, Dawson, Tokita, Aldering,
  Amanullah, Connolly, Doi, Faccioli, Fadeyev, Fruchter, {et~al.}}]{bar2009}
Barbary K. {et~al.}, 2009, ApJ, 690, 1358

\bibitem[{{Chen} {et~al}\mbox{.}(2013){Chen}, {Smartt}, {Bresolin},
  {Pastorello}, {Kudritzki}, {Kotak}, {McCrum}, {Fraser}, \&
  {Valenti}}]{chen2013}
{Chen} T.-W. {et~al.}, 2013, ApJ, 763, L28

\bibitem[{Chevalier \& Irwin(2011)}]{che2011}
Chevalier R.~A., Irwin C.~M., 2011, ApJ , 729, L6

\bibitem[{Chomiuk {et~al}\mbox{.}(2011)Chomiuk, Chornock, Soderberg, Berger,
  Chevalier, Foley, Huber, Narayan, Rest, Gezari, {et~al.}}]{chom2011}
Chomiuk L. {et~al.}, 2011, ApJ, 743, 114

\bibitem[{Cooke {et~al}\mbox{.}(2012)Cooke, Sullivan, Gal-Yam, Barton,
  Carlberg, Ryan-Weber, Horst, Omori, \& D{\'\i}az}]{coo2012}
Cooke J. {et~al.}, 2012, Nature, 491, 228

\bibitem[{Drake {et~al}\mbox{.}(2009)Drake, Djorgovski, Mahabal, Beshore,
  Larson, Graham, Williams, Christensen, Catelan, Boattini, {et~al.}}]{dra2009}
Drake A. {et~al.}, 2009, ApJ, 696, 870

\bibitem[{Gal-Yam(2012)}]{gal2012}
Gal-Yam A., 2012, Science, 337, 927

\bibitem[{Ginzburg \& Balberg(2012)}]{gin2012}
Ginzburg S., Balberg S., 2012, ApJ, 757, 178

\bibitem[{Howell {et~al}\mbox{.}(2013)Howell, Kasen, Lidman, Sullivan, Conley,
  Astier, Balland, Carlberg, Fouchez, Guy, {et~al.}}]{how2013}
Howell D. {et~al.}, 2013, ApJ, 779, 98

\bibitem[{Inserra {et~al}\mbox{.}(2013)Inserra, Smartt, Jerkstrand, Valenti,
  Fraser, Wright, Smith, Chen, Kotak, Pastorello, {et~al.}}]{ins2013}
Inserra C. {et~al.}, 2013, ApJ, 770, 128

\bibitem[{Kaiser {et~al}\mbox{.}(2010)Kaiser, Burgett, Chambers, Denneau,
  Heasley, Jedicke, Magnier, Morgan, Onaka, \& Tonry}]{kai2010}
Kaiser N. {et~al.}, 2010, in SPIE Astronomical Telescopes+ Instrumentation,
  International Society for Optics and Photonics, pp. 77330E--77330E

\bibitem[{Kasen \& Bildsten(2010)}]{kas2010}
Kasen D., Bildsten L., 2010, ApJ, 717, 245

\bibitem[{Kasen, Metzger \& Bildsten(2015)Kasen, Metzger, \&
  Bildsten}]{kas2015}
Kasen D., Metzger B., Bildsten L., 2015, ArXiv e-prints

\bibitem[{Leloudas {et~al}\mbox{.}(2012)Leloudas, Chatzopoulos, Dilday,
  Gorosabel, Vinko, Gallazzi, Wheeler, Bassett, Fischer, Frieman,
  {et~al.}}]{lel2012}
Leloudas G. {et~al.}, 2012, A\&A, 541, A129

\bibitem[{{Leloudas} {et~al}\mbox{.}(2015){Leloudas}, {Schulze}, {Kr{\"u}hler},
  {Gorosabel}, {Christensen}, {Mehner}, {de Ugarte Postigo}, {Amor{\'{\i}}n},
  {Th{\"o}ne}, {Anderson}, {Bauer}, {Gallazzi}, {He{\l}miniak}, {Hjorth},
  {Ibar}, {Malesani}, {Morell}, {Vinko}, \& {Wheeler}}]{lel2015}
{Leloudas} G. {et~al.}, 2015, MNRAS, 449, 917

\bibitem[{Lunnan {et~al}\mbox{.}(2014)Lunnan, Chornock, Berger, Laskar, Fong,
  Rest, Sanders, Challis, Drout, Foley, {et~al.}}]{lun2014}
Lunnan R. {et~al.}, 2014, ApJ, 787, 138

\bibitem[{Lunnan {et~al}\mbox{.}(2013)Lunnan, Chornock, Berger, Milisavljevic,
  Drout, Sanders, Challis, Czekala, Foley, Fong, {et~al.}}]{lun2013}
Lunnan R. {et~al.}, 2013, ApJ, 771, 97

\bibitem[{McCrum {et~al}\mbox{.}(2015)McCrum, Smartt, Rest, Smith, Kotak,
  Rodney, Young, Chornock, Berger, Foley, Fraser, Wright, Scolnic, Tonry,
  Urata, Huang, Pastorello, Botticella, Valenti, Mattila, Kankare, Farrow,
  Huber, Stubbs, Kirshner, Bresolin, Burgett, Chambers, Draper, Flewelling,
  Jedicke, Kaiser, Magnier, Metcalfe, Morgan, Price, Sweeney, Wainscoat, \&
  Waters}]{mcc2015}
McCrum M. {et~al.}, 2015, MNRAS,
  448, 1206

\bibitem[{{Moriya} \& {Maeda}(2012)}]{mor2012}
{Moriya} T.~J., {Maeda} K., 2012, ApJ , 756, L22

\bibitem[{{Nakar} \& {Piro}(2014)}]{nak2014}
{Nakar} E., {Piro} A.~L., 2014, ApJ, 788, 193

\bibitem[{Neill {et~al}\mbox{.}(2011)Neill, Sullivan, Gal-Yam, Quimby, Ofek,
  Wyder, Howell, Nugent, Seibert, Martin, {et~al.}}]{nei2011}
Neill J.~D. {et~al.}, 2011, ApJ, 727, 15

\bibitem[{Nicholl {et~al}\mbox{.}(2015{\natexlab{a}})Nicholl, Smartt,
  Jerkstrand, Sim, Inserra, Anderson, Baltay, Benetti, Chambers, Chen,
  {et~al.}}]{nic2015a}
Nicholl M. {et~al.}, 2015{\natexlab{a}}, ApJ ,
  807, L18

\bibitem[{Nicholl {et~al}\mbox{.}(2014)Nicholl, Smartt, Jerkstrand, Inserra,
  Anderson, Baltay, Benetti, Chen, Elias-Rosa, Feindt, Fraser, Gal-Yam,
  Hadjiyska, Howell, Kotak, Lawrence, Leloudas, Margheim, Mattila, McCrum,
  McKinnon, Mead, Nugent, Rabinowitz, Rest, Smith, Sollerman, Sullivan, Taddia,
  Valenti, Walker, \& Young}]{nic2014}
Nicholl M. {et~al.}, 2014, MNRAS,
  444, 2096

\bibitem[{Nicholl {et~al}\mbox{.}(2015{\natexlab{b}})Nicholl, Smartt,
  Jerkstrand, Inserra, Sim, Chen, Benetti, Fraser, Gal-Yam, Kankare, Maguire,
  Smith, Sullivan, Valenti, Young, Baltay, Bauer, Baumont, Bersier, Botticella,
  Childress, Dennefeld, Della~Valle, Elias-Rosa, Feindt, Galbany, Hadjiyska,
  Le~Guillou, Leloudas, Mazzali, McKinnon, Polshaw, Rabinowitz, Rostami,
  Scalzo, Schmidt, Schulze, Sollerman, Taddia, \& Yuan}]{nic2015b}
Nicholl M. {et~al.}, 2015{\natexlab{b}}, MNRAS, 452, 3869

\bibitem[{Ofek {et~al}\mbox{.}(2010)Ofek, Rabinak, Neill, Arcavi, Cenko,
  Waxman, Kulkarni, Gal-Yam, Nugent, Bildsten, {et~al.}}]{ofek2010}
Ofek E. {et~al.}, 2010, ApJ, 724, 1396

\bibitem[{Papadopoulos {et~al}\mbox{.}(2015)Papadopoulos, D'Andrea, Sullivan,
  Nichol, Barbary, Biswas, Brown, Covarrubias, Finley, Fischer,
  {et~al.}}]{pap2015}
Papadopoulos A. {et~al.}, 2015, MNRAS, 449, 1215

\bibitem[{Piro(2015)}]{piro2015}
Piro A.~L., 2015, ApJ , 808, L51

\bibitem[{Quimby {et~al}\mbox{.}(2011)Quimby, Kulkarni, Kasliwal, Gal-Yam,
  Arcavi, Sullivan, Nugent, Thomas, Howell, Nakar, {et~al.}}]{qui2011}
Quimby R.~M. {et~al.}, 2011, Nature, 474, 487

\bibitem[{Quimby {et~al}\mbox{.}(2013)Quimby, Yuan, Akerlof, \&
  Wheeler}]{qui2013}
Quimby R.~M., Yuan F., Akerlof C., Wheeler J.~C., 2013, MNRAS, 431, 912

\bibitem[{Rabinak \& Waxman(2011)}]{rab2011}
Rabinak I., Waxman E., 2011, ApJ, 728, 63

\bibitem[{Rau {et~al}\mbox{.}(2009)Rau, Kulkarni, Law, Bloom, Ciardi,
  Djorgovski, Fox, Gal-Yam, Grillmair, Kasliwal, {et~al.}}]{rau2009}
Rau A. {et~al.}, 2009, PASP,
  121, 1334

\bibitem[{Smartt {et~al}\mbox{.}(2015)Smartt, Valenti, Fraser, Inserra, Young,
  Sullivan, Pastorello, Benetti, Gal-Yam, {Knapic, C.}, {Molinaro, M.},
  {Smareglia, R.}, {Smith, K. W.}, {Taubenberger, S.}, {Yaron, O.}, {Anderson,
  J. P.}, {Ashall, C.}, {Balland, C.}, {Baltay, C.}, {Barbarino, C.}, {Bauer,
  F. E.}, {Baumont, S.}, {Bersier, D.}, {Blagorodnova, N.}, {Bongard, S.},
  {Botticella, M. T.}, {Bufano, F.}, {Bulla, M.}, {Cappellaro, E.}, {Campbell,
  H.}, {Cellier-Holzem, F.}, {Chen, T.-W.}, {Childress, M. J.}, {Clocchiatti,
  A.}, {Contreras, C.}, {Dall’Ora, M.}, {Danziger, J.}, {de Jaeger, T.}, {De
  Cia, A.}, {Della Valle, M.}, {Dennefeld, M.}, {Elias-Rosa, N.}, {Elman, N.},
  {Feindt, U.}, {Fleury, M.}, {Gall, E.}, {Gonzalez-Gaitan, S.}, {Galbany, L.},
  {Morales Garoffolo, A.}, {Greggio, L.}, {Guillou, L. L.}, {Hachinger, S.},
  {Hadjiyska, E.}, {Hage, P. E.}, {Hillebrandt, W.}, {Hodgkin, S.}, {Hsiao, E.
  Y.}, {James, P. A.}, {Jerkstrand, A.}, {Kangas, T.}, {Kankare, E.}, {Kotak,
  R.}, {Kromer, M.}, {Kuncarayakti, H.}, {Leloudas, G.}, {Lundqvist, P.},
  {Lyman, J. D.}, {Hook, I. M.}, {Maguire, K.}, {Manulis, I.}, {Margheim, S.
  J.}, {Mattila, S.}, {Maund, J. R.}, {Mazzali, P. A.}, {McCrum, M.},
  {McKinnon, R.}, {Moreno-Raya, M. E.}, {Nicholl, M.}, {Nugent, P.}, {Pain,
  R.}, {Pignata, G.}, {Phillips, M. M.}, {Polshaw, J.}, {Pumo, M. L.},
  {Rabinowitz, D.}, {Reilly, E.}, {Romero-Cañizales, C.}, {Scalzo, R.},
  {Schmidt, B.}, {Schulze, S.}, {Sim, S.}, {Sollerman, J.}, {Taddia, F.},
  {Tartaglia, L.}, {Terreran, G.}, {Tomasella, L.}, {Turatto, M.}, {Walker,
  E.}, {Walton, N. A.}, {Wyrzykowski, L.}, {Yuan, F.}, \& {Zampieri,
  L.}}]{sma2015}
Smartt S.~J. {et~al.}, 2015, A\&A, 579, A40

\bibitem[{Vreeswijk {et~al}\mbox{.}(2014)Vreeswijk, Savaglio, Gal-Yam, De~Cia,
  Quimby, Sullivan, Cenko, Perley, Filippenko, Clubb, {et~al.}}]{vre2014}
Vreeswijk P.~M. {et~al.}, 2014, ApJ, 797, 24

\bibitem[{Woosley(2010)}]{woo2010}
Woosley S., 2010, ApJ , 719, L204

\end{thebibliography}

\end{document}